\documentclass[prd,aps,tightenlines,twocolumn,nofootinbib,preprintnumbers]{revtex4}
\usepackage{amsmath,amssymb}
\makeatother

\def\un#1{\relax\ifmmode\@@underline#1\else
        $\@@underline{\hbox{#1}}$\relax\fi}

\let\du=\du                     

\def\d{\delta}

\def\h{\eta}

\def\j{\psi}

\def\p{\pi}
\def\q{\theta}

\def\s{\sigma}

\def\F{\Phi}

\def\ce{{\cal E}}

\def\car{{\cal R}}

\def\bo{{\raise-.3ex\hbox{\large$\Box$}}}               
\def\TH{{\raise.2ex\hbox{$\displaystyle \bigodot$}\mskip-4.7mu \llap H \;}}
\def\face{{\raise.2ex\hbox{$\displaystyle \bigodot$}\mskip-2.2mu \llap {$\ddot
        \smile$}}}                                      

\def\Bar#1{\overline{#1}}                       
\def\abs#1{\left| #1\right|}                    
\def\leftrightarrowfill{$\mathsurround=0pt \mathord\leftarrow \mkern-6mu
        \cleaders\hbox{$\mkern-2mu \mathord- \mkern-2mu$}\hfill
        \mkern-6mu \mathord\rightarrow$}
\def\dvec#1{\vbox{\ialign{##\crcr
        \leftrightarrowfill\crcr\noalign{\kern-1pt\nointerlineskip}
        $\hfil\displaystyle{#1}\hfil$\crcr}}}           

\def\frac#1#2{{\textstyle{#1\over\vphantom2\smash{\raise.20ex
        \hbox{$\scriptstyle{#2}$}}}}}                   
\def\sfrac#1#2{{\vphantom1\smash{\lower.5ex\hbox{\small$#1$}}\over
        \vphantom1\smash{\raise.4ex\hbox{\small$#2$}}}} 
\def\bfrac#1#2{{\vphantom1\smash{\lower.5ex\hbox{$#1$}}\over
        \vphantom1\smash{\raise.3ex\hbox{$#2$}}}}       
\def\afrac#1#2{{\vphantom1\smash{\lower.5ex\hbox{$#1$}}\over#2}}    

\def\[{\lfloor{\hskip 0.35pt}\!\!\!\lceil}
\def\]{\rfloor{\hskip 0.35pt}\!\!\!\rceil}

\def\du#1#2{_{#1}{}^{#2}}

\def\fracm#1#2{\hbox{\large{${\frac{{#1}}{{#2}}}$}}}

\def\un{\underline}
\def\fracmm#1#2{{{#1}\over{#2}}}

\def\low#1{{\raise -3pt\hbox{${\hskip 0.75pt}\!_{#1}$}}}

\newskip\humongous \humongous=0pt plus 1000pt minus 1000pt
\def\caja{\mathsurround=0pt}
\def\eqalign#1{\,\vcenter{\openup2\jot \caja
        \ialign{\strut \hfil$\displaystyle{##}$&$
        \displaystyle{{}##}$\hfil\crcr#1\crcr}}\,}
\newif\ifdtup

\newcommand{\be}{\begin{equation}}
\newcommand{\ee}{\end{equation}}
\newcommand{\nbe}{\begin{equation*}}
\newcommand{\nee}{\end{equation*}}

\newcommand{\lb}{\label}

\begin{document}

\title{Embedding $(R+R^2)$-Inflation into Supergravity}
\author{Sergei V. Ketov~${}^{a,b}$ and Alexei A. Starobinsky~${}^{c,d}$} 

\affiliation{${}^a$~Department of Physics, Tokyo Metropolitan University, 
Minami-ohsawa 1-1, Hachioji-shi, Tokyo 192-0397, Japan \\
${}^b$~Institute for the Physics and Mathematics of the Universe (IPMU), The 
University of Tokyo, Kashiwanoha 5-1-5, Kashiwa-shi, Chiba 277-8568, Japan \\
${}^c$~Landau Institute for Theoretical Physics, Moscow 119334, Russia\\
${}^d$~Research Center for the Early Universe (RESCEU), Graduate School of Science,
The University of Tokyo, Tokyo 113-0033, Japan}

\begin{abstract}
We find the natural embedding of the $(R+R^2)$-inflationary model into the recently 
constructed $F(\car)$ supergravity. It gives a simple and viable realization of 
chaotic inflation in supergravity. The only requirement for a slow-roll inflation is
 the existence of the $\car^3$-term with an anomalously large coefficient in Taylor 
expansion of the $F(\car)$-function.

\end{abstract}

\preprint{RESCEU--24/10}

\maketitle


\section{Introduction}

A natural realization of inflation in supergravity is known to be problematic 
\cite{jap1,kl} because of the factor $\exp(K/M^2_{\rm Pl})$ in the (F-term 
generated) scalar potential \cite{crem}, where $K$ is the K\"ahler potential
of the chiral scalar matter superfields $\F$ and $\Bar{\F}$. The naive (tree-level)
{\it Ansatz} $K=\bar{\F}\F$ gives rise to the scalar potential proportional to
$\exp(\Bar{\F}\F)$ that is too steep for a slow-roll inflation (the so-called 
$\h$-problem) with the unacceptable inflaton mass 
$\abs{m^2}\sim V_0/M^2_{\rm Pl}\approx H^2$. 

To cure the above problem, the D-term mechanism was proposed \cite{bdva}, where 
the inflation is generated in the {\it gauge} sector and is highly sensitive to
the gauge charges. Another proposal is to assume that the K\"ahler potential does
not depend upon some scalars (= {\it flat directions}) and then add a desired 
scalar super-potential for the flat directions \cite{jap2}. Both proposals are 
non-geometrical and non-universal because they refer to the matter sector (not 
gravity) and require the existence of extra fields too.

As is also known for a long time  \cite{star,star2}, viable inflationary models can 
be easily constructed in (non-supersymmetric) $f(R)$-gravity theories (see eg., 
refs.~\cite{fgrev} for a recent review) with the action
\be \lb{fgr} S=-\fracmm{M^2_{\rm Pl}}{2} \int d^4x \sqrt{-g}~ f(R)
\ee
whose function $f(R)$ begins with the scalar curvature $R$, and the difference 
$(f(R)-R)$ takes the form $R^2A(R)$ for $R\to\infty$, with a slowly varying function 
$A(R)$ (we assume that $\hbar=c=1$). The simplest one of those models is given  by 
(see ref.~\cite{kan} for our sign conventions) 
\be \lb{star}
f(R) = R-\fracmm{R^2}{6M^2} 
\ee
The theory (\ref{star}) is known as the excellent model of chaotic inflation 
\cite{linde}. The coefficient in front of the second term on the right-hand-side of 
eq.~(\ref{star}) is chosen so that $M$ actually coincides with the rest mass of the 
scalar particle appearing in $f(R)$-gravity (dubbed {\it scalaron} in 
ref.~\cite{star}) at low curvatures $\abs{R}\ll M^2$ or in flat spacetime, 
in particular. The model fits the observed 
amplitude of scalar perturbations if  $M/M_{\rm Pl} \approx 1.5 \cdot  10^{-5} 
(50/N_e)$, and gives rise to the spectral index  
$n_s-1\approx -2/N_e\approx -0.04(50/N_e)$ 
and the scalar-to-tensor ratio $r\approx 12/N_e^2 \approx 0.005(50/N_e)^2$, 
in terms of the e-foldings number $N_{e}\approx (50\div 55)$ depending upon details 
of reheating after inflation \cite{mukh,kan}. Despite of the fact that it is known 
for 30 years, the model (\ref{star}) remains viable and is in agreement with the 
most recent WMAP7  observations of $n_s=0.963\pm 0.012$ and $r<0.24$  (with 95\% CL)
 \cite{wmap}.

The purpose of this Letter is to show that there exists a natural embedding of 
the inflationary model (\ref{star}) into supergravity.~\footnote{For completeness, 
it is worthwhile to mention some other microscopic approaches that are unrelated to 
supergravity but also lead to the $(R+R^2)$-model as the macroscopic (and 
approximate) theory with a high precision : (i) the Higgs inflation with a large 
non-minimal coupling of the Higgs field to gravity \cite{bez,bks}, and (ii) the 
so-called emergent gravity \cite{kvol}.}  For that purpose we use the supersymmetric
 extension of $f(R)$ gravity theories, called $F(\car)$ supergravity that was 
recently constructed in ref.~\cite{us}. In Sec.~2 we briefly outline the $F(\car)$ 
supergravity by focusing on its reduction to the more familiar $f(R)$ gravity. In 
Sec.~3 we propose a simple realization of chaotic inflation in supergravity via 
embedding of the bosonic model (\ref{star}) into the particular $F(\car)$ 
supergravity model. Sec.~4 is our conclusion.  

\section{$F(\car)$ supergravity and $f(R)$ gravity}

The most succinct formulation of $F(\car)$ supergravity exist in a chiral
4D, $N=1$ superspace where  it is defined by an action~\footnote{For simplicity,
we take $M_{\rm Pl}=1$ in this section.}
\be  \lb{act}
 S = \int d^4x d^2\q\, \ce F(\car) + {\rm H.c.}
\ee
in terms of a holomorphic function $F(\car)$ of the covariantly-chiral scalar
curvature superfield $\car$, and the chiral superspace density $\ce$. The chiral
$N=1$ superfield $\car$ has the scalar curvature $R$ as the field coefficient
at its $\q^2$-term (see eg., ref.~\cite{sspace} for details about supergravity
in superspace). The chiral superspace density $\ce$ (in a WZ gauge) reads
\be \lb{cde}
\ce = e \left( 1- 2i\q\s_a\bar{\j}^a +\q^2 B\right) 
\ee
where $e=\sqrt{-g}$, $\j^a$ is gravitino, and $B=S-iP$ is the complex scalar 
auxiliary field (it does not propagate in the theory (\ref{act}) despite of the
apparent presence of the higher derivatives). The full component structure of the 
action (\ref{act}) is very complicated. Nevertheless, it is classically equivalent
 to the standard $N=1$ Poincar\'e supergravity minimally coupled to the chiral 
scalar superfield, via the supersymmetric Legendre-Weyl-K\"ahler transform 
\cite{us}. The chiral scalar superfield is given by the superconformal mode of the
supervielbein (in Minkowski or AdS vacuum) which becomes dynamical in $F(\car)$ 
supergravity.   

A relation to the $f(R)$-gravity theories can be established by dropping the
gravitino $(\j^a=0)$ and restricting the auxiliary field $B$ to its real (scalar)
component, $B=3X$ with $\Bar{X}=X$. Then, as was shown in ref.~\cite{fgra}, the
bosonic Lagrangian takes the form
\be \lb{bos}
L = 2F' \left[ \frac{1}{3}R +4X^2 \right] + 6XF
\ee
It follows that the auxiliary field $X$ obeys an algebraic equation of motion,
\be \lb{aux}
3F +  11F'X + F''\left[ \frac{1}{3}R + 4X^2\right] =0
\ee
In those equations $F=F(X)$ and the primes denote the derivatives with respect to
$X$. Solving eq.~(\ref{aux}) for $X$ and substituting the solution back into 
eq. (\ref{bos}) results in the bosonic function $L=-\frac{1}{2}f(R)$. 

It is natural to expand the input function $F(\car)$ into power series of $\car$.
For instance, when $F(\car)=f_0 -\frac{1}{2}f_1\car$ with some (non-vanishing and 
complex) coefficients $f_0$ and $f_1$, one recoveres the standard {\it pure} $N=1$ 
supergrvity with a negative cosmological term \cite{us}. 

A more interesting  Ansatz is given by
\be \lb{qua} 
F(\car)= -\frac{1}{2}f_1 \car + \frac{1}{2}f_2 \car^2
\ee 
with some real coefficients $f_1$ and $f_2$. It gives rise to the bosonic function
(with $f_1=3/2$) \cite{fgra}
\be \lb{bos2}
\eqalign{ 
f(R) & ~= \fracmm{5\cdot 17}{3^2\cdot 11}R -
\fracmm{2^2\cdot 7}{3^2\cdot 11}(R-R_{\rm max})\left[ 1-\sqrt{1-R/R_{\rm max}}\right] 
\cr 
& ~= R - \fracmm{R^2}{6M^2}  -\fracmm{11R^3}{252M^4}  + {\cal O}(R^4)  \cr} 
\ee
where $R_{\rm max}=\fracmm{3^2\cdot 7^2}{2^3\cdot 11}f^{-2}_2$ is the AdS bound 
automatically generated in the model, and  $M^2=\fracmm{11}{7}R_{\rm max}$.
Unfortunately, the model (\ref{bos2}) is not viable as the inflationary model because 
it suffers from the $\h$-problem arising due to the presence of the higher-order terms 
with respect to the scalar curvature in eq.~(\ref{bos2}) \cite{fgra}.

\section{Our new model}

The Ansatz we propose in this Letter is given by 
\be \lb{cub}
F(\car)= -\frac{1}{2}f_1 \car + \frac{1}{2}f_2 \car^2 -\frac{1}{6}f_3\car^3
\ee
whose real (positive) coupling constants $f_{1,2,3}$ are of (mass) dimension $2$, $1$
and $0$, respectively.  Our conditions on the coefficients are
\be \lb{cco}
 f_3 \gg 1~,\qquad f_2^2 \gg f_1  
\ee 
The first condition is needed to have inflation at the curvatures much less than
$M^2_{\rm Pl}$ (and to meet observations), while the second condition is needed to 
have the scalaron (inflaton) mass be much less than $M_{\rm Pl}$, in order to avoid 
large (gravitational) quantum loop corrections  after the end of inflation up to the
present time.


Stability of our bosonic embedding (\ref{bos}) in supergravity implies 
$F'(X)<0$. In the case (\ref{cub}) it gives rise to the condition $f_2^2 < f_1f_3$. 
For simplicity we will  assume 
\be \lb{adcon}
f_2^2\ll f_1f_3 
\ee
Then the second term on the right-hand-side of eq.~(\ref{cub}) will not affect 
inflation, as is shown below. 


Equation (\ref{bos}) with the Ansatz (\ref{cub}) reads
\be \lb{bos3}
L = -5f_3X^4 + 11f_2 X^3 - (7f_1 +\frac{1}{3}f_3R)X^2 +\frac{2}{3}f_2RX 
-\frac{1}{3}f_1R
\ee 
and gives rise to a cubic equation on $X$,
\be \lb{aux3} 
X^3 -\left( \fracmm{33f_2}{20f_3}\right)X^2 +\left( \fracmm{7f_1}{10f_3} 
+\fracmm{1}{30}R\right)X - \fracmm{f_2}{30f_3}R =0
\ee
We find three consecutive (overlapping) regimes.
\begin{itemize}
\item The high curvature regime including inflation  is given by 
\be \lb{reg1}
\d R<0 \quad {\rm and} \quad 
\fracmm{\abs{\d R}}{R_0}\gg \left(\fracmm{f^2_2}{f_1f_3}\right)^{1/3}
\ee
where we have introduced the notation $R_0=21f_1/f_3>0$ and $\d R = R+R_0$. With our sign 
conventions (Sec.~I) we have $R<0$ during the de Sitter and matter dominated stages. 
In the regime (\ref{reg1})  the $f_2$-dependent terms in eqs.~(\ref{bos3}) and 
(\ref{aux3}) can be neglected, and we get
\be \lb{aux31}
X^2 = -\frac{1}{30} \d R
\ee
and
\be \lb{lag1}
L = -\fracmm{f_1}{3}R + \fracmm{f_3}{180}(R+R_0)^2
\ee
It closely reproduces the inflationary model (\ref{star}) since inflation occurs at 
$\abs{R}\gg R_0$. So it is natural to denote $f_3=15M^2_{\rm Pl}/M^2$ (see Sec.~I).
It is worth mentioning that we cannot simply set $f_2=0$ in 
eq.~(\ref{cub}) because it would imply $X=0$ and $L=-\frac{f_1}{3}R$ for $\d R>0$. 
As a result of that the scalar degree of freedom would disappear that would lead to 
the breaking of a regular Cauchy evolution. Therefore, the second term in 
eq.~(\ref{cub}) is needed to remove that degeneracy.
\item The intermediate (post-inflationary) regime is given by
\be \lb{reg2} 
 \fracmm{\abs{\d R}}{R_0}\ll 1
\ee
In this case $X$ is given by a root of the cubic equation
\be \lb{root2}
 30 X^3 +(\d R)X +\fracmm{f_2R_0}{f_3}=0
\ee
It also implies that the 2nd term in eq.~(\ref{aux3}) is always small.
Equation (\ref{root2}) reduces to eq.~(\ref{aux31}) under the conditions 
(\ref{reg1}).
\item The low-curvature regime (up to $R=0$) is given by
\be \lb{reg3}
\d R>0 \quad {\rm and} \quad 
\fracmm{\d R}{R_0}\gg \left(\fracmm{f^2_2}{f_1f_3}\right)^{1/3}
\ee
It yields
\be \lb{alow}
X = \fracmm{f_2R}{f_3(R+R_0)}
\ee
and
\be \lb{lag3}
L = -\fracmm{f_1}{3}R + \fracmm{f_2^2R^2}{3f_3(R+R_0)} 
\ee
It is now clear that $f_1$ should be equal to $3M_{Pl}^2/2$
in order to obtain the correctly normalized Einstein gravity 
at $|R|\ll R_0$. In this regime the scalaron mass squared 
 is given by
\be \lb{smass}
 \fracmm{1}{3\abs{f''(R)}}=\fracmm{f_3R_0M_{\rm Pl}^2}{4f_2^2}
= \fracmm{21f_1}{4f_2^2}M_{\rm Pl}^2 = \fracmm{63M_{\rm Pl}^4}{8f_2^2}
\ee
in agreement with the case of the absence of the ${\cal R}^3$ term,
studied in the previous section. The scalaron mass squared (\ref{smass}) is
much less than $M_{Pl}^2$ indeed, due to the second  inequality in eq.~(\ref{cco}),
but it is much more than the one  at the end of inflation $(\sim M^2)$.
\end{itemize}

It is worth noticing that the corrections to the Einstein action in
eqs.~(\ref{lag1}) and (\ref{lag3})  are of ther same order (and small) 
at the borders of the intermediate region (\ref{reg2}).

The roots of the cubic equation (\ref{aux3}) are given by the textbook (Cardano)
 formula \cite{abs}, though that formula is not very illuminating in a generic case. 
The Cardano formula greatly simplifies in the most interesting (high curvature)
regime where inflation takes place, and the Cardano discriminant is
\be \lb{disc}
D \approx \left( \fracm{R}{90}\right)^3 < 0
\ee
It implies that all three roots are real and unequal. The Cardano formula yields the 
roots
\be \lb{card}
X_{1,2,3}\approx \fracmm{2}{3}\sqrt{ \fracmm{-R}{10} } \cos
\left( \fracmm{27}{4f_3 \sqrt{-10R/f^2_2}} + C_{1,2,3} \right) +\fracmm{11f_2}{20f_3}
\ee 
where the constant $C_{1,2,3}$ takes the values $(\p/6,5\p/6,3\p/2)$. 

As regards the leading terms, eqs.~(\ref{bos3}) and (\ref{card}) result in the 
$(-R)^{3/2}$ correction to the $(R+R^2)$-terms in the effective Lagrangian in the 
high-curvature regime $|R|\gg f_2^2/f_3^2$. In order to verify that this correction 
does not change our results under the conditions 
(\ref{reg1}), let's consider the $f(R)$-gravity model with
\be  \lb{try}  f(R) = R - b(-R)^{3/2} - aR^2 \ee
whose parameters $a>0$ and $b>0$ are subject to the conditions $a\gg 1$ and  
$b/a^2\ll 1$. It is easy to check that $f'(R)>0$ for $R\in (-\infty,0]$, as is needed
for (classical) stability.

Any $f(R)$ gravity model is known to be classiclally equivalent to the scalar-tensor 
gravity with  proper scalar potential \cite{equiv}. The scalar potential can be 
calculated from a given function $f(R)$ along the standard lines (see eg., 
refs.~\cite{fgrev,kan}). We find (in the high curvature regime)
\be \lb{poten}
V(y) = \fracmm{1}{8a} \left( 1- e^{-y}\right)^2 +
\fracmm{b}{8\sqrt{2a}}e^{-2y}\left(e^y-1\right)^{3/2}
\ee
in terms of the inflaton field $y$. The first term of this equation is the scalar 
potential associated with the pure $(R+R^2)$ model, and the 2nd term is the 
correction due to the $R^{3/2}$-term in eq.~(\ref{try}). It is now clear that for 
large positive $y$ the vacuum energy in  the first term dominates and drives 
inflation until the vacuum energy is compensated by the $y$-dependent terms near 
$e^y=1$.   

It can be verified along the lines of ref.~\cite{mukh} that 
the formula for scalar perturbations remains the same as for
the model (\ref{star}), ie. $\Delta_{\cal R}^2\approx 
N^2M^2/(24\pi^2M_{\rm Pl}^2)$, where $N$ is the number of e-folds 
from the end of inflation. So, to fit the observational data, one 
has to choose $f_3 \approx 5N_e^2/(8\pi^2 \Delta_{\cal R}^2)\approx
6.5\cdot 10^{10} (N_e/50)^2$. Here the value of $\Delta_{\cal R}$ is 
taken from ref.~\cite{wmap} and the subscript ${\cal R}$ has a different 
meaning from the rest of this paper.

\section{Conclusion}

We conclude that the model (\ref{cub}) with a sufficiently small $f_2$ obeying the
conditions (\ref{cco}) and (\ref{adcon}) gives a viable realization of the chaotic 
$(R+R^2)$-type inflation 
in supergravity. The only significant difference with respect to the original $(R+R^2)$
inflationary model is the scalaron mass that becomes much larger than $M$ in supergravity,
soon after the end of inflation when $\d R$ becomes positive. However, it only makes the 
scalaron decay faster and creation of the usual matter (reheating) more effective.

The whole series in powers of ${\car}$ may also be considered, instead of the limited
Ansatz (\ref{cub}). The only necessary condition for embedding inflation is that $f_3$
should be anomalously large. When the curvature grows, the $\car^3$-term should become 
important much earlier than the convergence radius of the whole series without that term. 
Of course, it means that viable inflation may not occur for any function $F(\car)$
 but only inside a small region of non-zero measure in the space of all those functions. 
However, the same is true for all known inflationary models, so the very existence of 
inflation has to be taken from the observational data, not from a pure thought.

We consider our results as the viable alternative to the earlier fundamental proposals
 \cite{bdva,jap2} for realization of chaotic inflation in supergravity. But inflation 
is not the only target of our construction. As is well known  \cite{star,star2} ---
see also the recent paper \cite{gorb} --- the scalaron decays into pairs of particles and 
anti-particles of quantum matter fields, while its decay into gravitons is strongly 
suppressed \cite{s81}. It represents the universal 
mechanism of viable reheating after inflation and provides a transition to the subsequent 
hot radiation-dominated stage of the Universe evolution and the characteristic temperature 
$T_{\rm reheating}\approx 10^9~GeV$. In its turn, it
leads to the standard primordial nucleosynthesis after. In $F(R)$ supergravity the 
scalaron has a pseudo-scalar superpartner (or axion) that may be the source of a strong 
CP-violation and then, subsequently, a leptogenesis and a baryogenesis that naturally 
lead to baryon (matter-antimatter) asymmetry \cite{fy}.

Supersymmetry in $F(R)$ supergravity is already broken by inflation. It may give rise
to a massive gravitino with $m_{3/2} \geq 10^7~GeV$. The gravitino is a natural candidate 
for the cold dark matter in our construction, {\it cf.} ref.~\cite{buchm}. The 
gravitationally mediated supersymmetry breaking may serve as the important element for 
the new particle phenomenology (beyond the Standard Model) based on a matter-coupled 
$F(R)$ supergravity.
 
\section*{Acknowledgements}

SK was supported in part by the Japanese Society for Promotion of Science (JSPS) and
the German Max-Planck-Society (Werner-Heisenberg Institute of Physics in Munich). 
AS acknowledges the RESCEU hospitality as a visiting professor. He was also partially 
supported by the Russian Foundation for Basic Research (RFBR) under the grant 
09--02--12417--ofi--m.

\end{document}